\documentstyle[mprocl]{article}

\input epsf
\def\Journal#1#2#3#4{{#1} {\bf #2}, #3 (#4)}

\def\NPB{{\em Nucl. Phys.} B}
\def\PLB{{\em Phys. Lett.}  B}
\def\PRL{\em Phys. Rev. Lett.}
\def\PRD{{\em Phys. Rev.} D}

\def\NPPS{{\em Nucl. Phys. Proc. Suppl.}}

\def\be{\begin{equation}}
\def\ee{\end{equation}}
\def\bea{\begin{eqnarray}}
\def\eea{\end{eqnarray}}

\begin{document}

\title{SUSY Searches at Tevatron Collider}

 \author{C. Pagliarone}

\address{University of Cassino \& INFN of Pisa,\\ 
via di Biasio,43 - 03043 Cassino, Italy\\E-mail: pagliarone@fnal.gov}   
%
\maketitle\abstracts{This article presents recent results 
of searches for Supersymmetry using the CDF and the D$\not$O
detector at the Fermilab Tevatron Collider. 
Described are the Tevatron searches for third generation 
scalar quarks and for supersymmetric signatures 
involving photons.
All the  reported results have been obtained assuming 
theoretical models in  which $\mathcal{R}$-parity is conserved.} 

\section{Introduction}

\subsection{Supersymmetry}\label{subsec:susy}
 
Although at present, the Standard Model (SM) 
provides a remarkably successful description of 
known phenomena, it seems to be, most likely, 
a low energy effective theory of spin-1/2 
matter fermions interacting via spin-1 
gauge bosons~\cite{Altarelli}.
An excellent candidate of a new theory, able to 
describe physics at arbitrarily high energies, 
is Supersymmetry (SUSY)~\cite{SUSYTEO}.
SUSY is a larger space-time symmetry, 
that relates bosons to fermions.
Even if we don't have direct 
experimental evidences of SUSY,
there are remarkable theoretical properties that provides 
ample motivation for its study.
SUSY describe electroweak data equally well than SM 
but, in addition, allows the unification of the 
gauge couplings constants, the unification of the 
Yukawa couplings and 
do not require the incredible fine tuning, 
endemic to Higgs sector of the SM.
Naturally, SUSY cannot be an exact symmetry 
of the nature, as none of the predicted 
spin 0 partners of the quarks or leptons 
and none of the spin 1/2 partners 
of the gauge bosons have been 
observed so far~\cite{BOER}.

\noindent
In Supersymmetry fermions can 
couple to a sfermion and a fermion,
violating lepton and/or baryon number.
To avoid this problem, a new quantum number,
the $\mathcal{R}$-parity, has been introduced~\cite{DREINER}. 
For SM particles ${\mathcal{R}}=\,+1$, for the SUSY 
partners ${\mathcal{R}}=\,-1$.
$\mathcal{R}$-parity conservation
has deep phenomenological consequences~\cite{VISSANI}. 
SUSY particles can be only produced in pairs; 
the lightest supersymmetric particle (LSP) does exist
and it is stable and interacts very weakly with the 
ordinary matter, leading to a robust 
missing transverse energy signature ($\not\!\!\!E_{\rm T}$);
the LSP is a natural candidate for the dark matter.
In the present article, we present a review of 
recent Tevatron searches performed assuming 
$\mathcal{R}$-parity conservation.

\subsection{Gauge Mediated SUSY Breaking Models}\label{subsec:gmsbm}

\indent
Theories with gauge-mediated supersymmetry
breaking  (GMSB) provide an interesting
alternative scenario.
In GMSB~\cite{GMSB1,GMSB2}, 
the dynamical supersymmetry breaking (DSB)
is mediated by gauge interactions.
In recent years, many mechanism for 
DSB have been found and realistic 
models have been constructed.
This class of models assumes that supersymmetry is broken with a scale 
$\sqrt{F}$ in a sector of the theory which contains heavy
non-Standard-Model particles. This sector then couples to a 
set of particles with Standard Model interactions, called 
messengers, which have a mass of order M. The mass splitting,
between the superpartners in the messenger multiplets, depends
by $\sqrt{F}$ and the SUSY particles get their masses 
via gauge interactions, so there are no flavor changing 
neutral currents. 
These theories have a very distinctive phenomenological 
features. The typical SUSY spectra is different 
from those in the SUGRA models; the LSP is the 
gravitino $\tilde{G}$ (in SUGRA 
$M_{\tilde{G}} \sim 1$ $TeV$),  the next 
lightest supersymmetric particle (NLSP) has a lifetime
that can vary strongly from model to model 
($1\mu<c\tau<$ several Km) and decays into $\tilde{G}$.

\section{Search for scalar top}

Search for scalar top is particularly 
interesting since, in many SUSY models, 
the top-squark  eigenstate 
$\tilde{t}_{1}$ (stop)  is expected to be the 
lightest squark~\cite{GENSTOP}.
The strong Yukawa coupling, between top/stop
and Higgs fields, gives rise, in fact, to potentially 
large mixing effects and mass splitting.
The CDF experiment has recently performed
two different searches
for scalar top: direct
$\bar{\tilde{t}}_{1}\tilde{t}_{1}$ 
production, assuming a branching ratio
${\mathcal{BR}}(\tilde{t}_{1}\rightarrow c\tilde{\chi}^{0}_{1})=\,\,100\%$, 
and a search for supersymmetric decays of the top quark: 
$t\rightarrow \tilde{t}_{1}\tilde{\chi}^{0}_{1}$, with 
$\tilde{t}_{1} \rightarrow b\tilde{\chi}^{\pm}_{1}$.

\begin{figure}[t!]
\begin{minipage}{2.4in}
  \epsfxsize2.4in 
\vspace{-0.5cm}
  \hspace*{-0.1cm}\epsffile{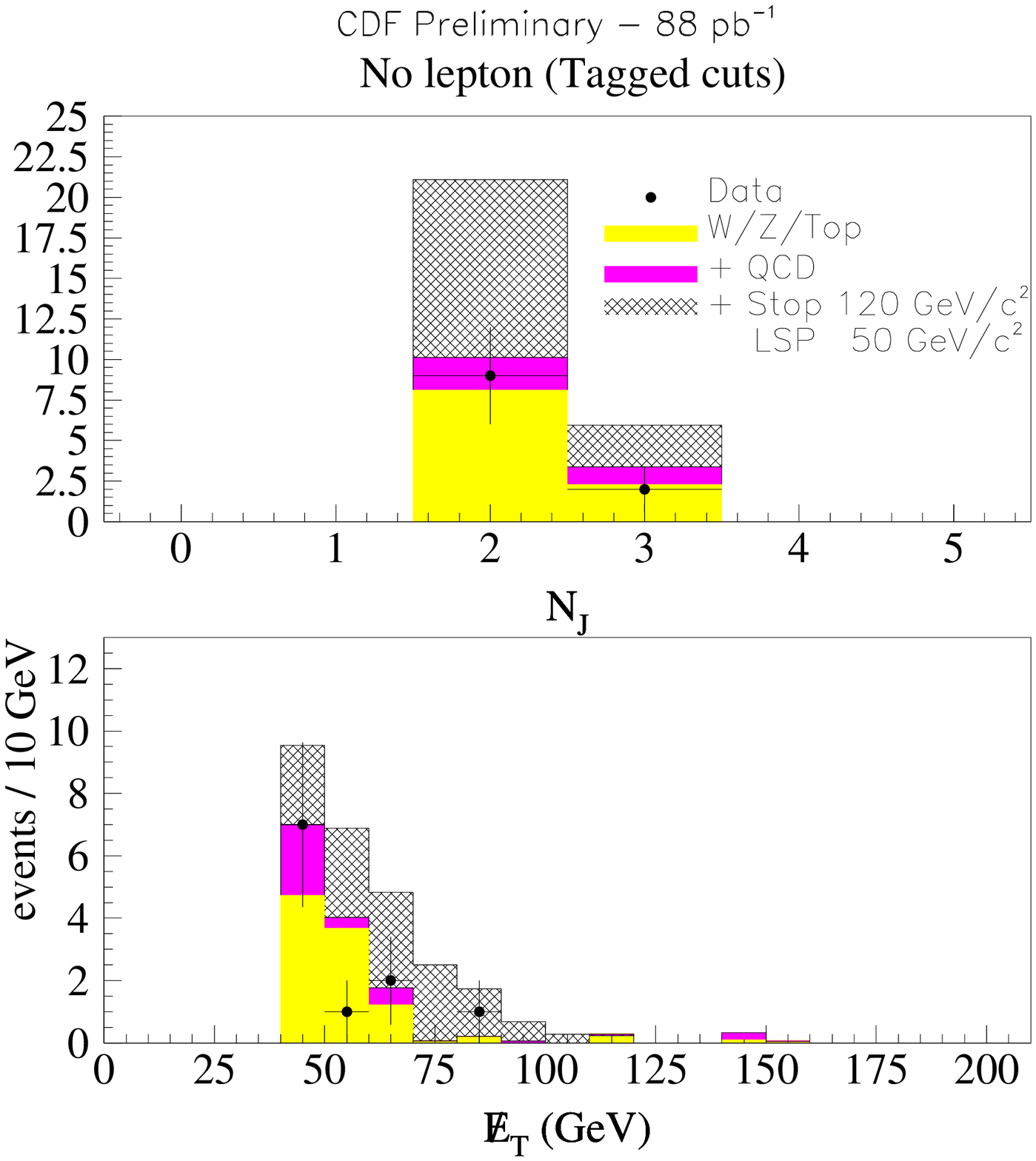}
\vspace{-1.25cm}
  \hspace*{-0.2cm}
\caption{\it Jet multiplicity and $\not\!\!\!E_{\rm T}$ 
distribution after all cuts have been 
applied (CDF experiment).}
\label{dirstop}
\end{minipage}\hfill
\begin{minipage}{2.4in}
  \epsfxsize2.4in 
  \hspace*{0.12cm}\epsffile{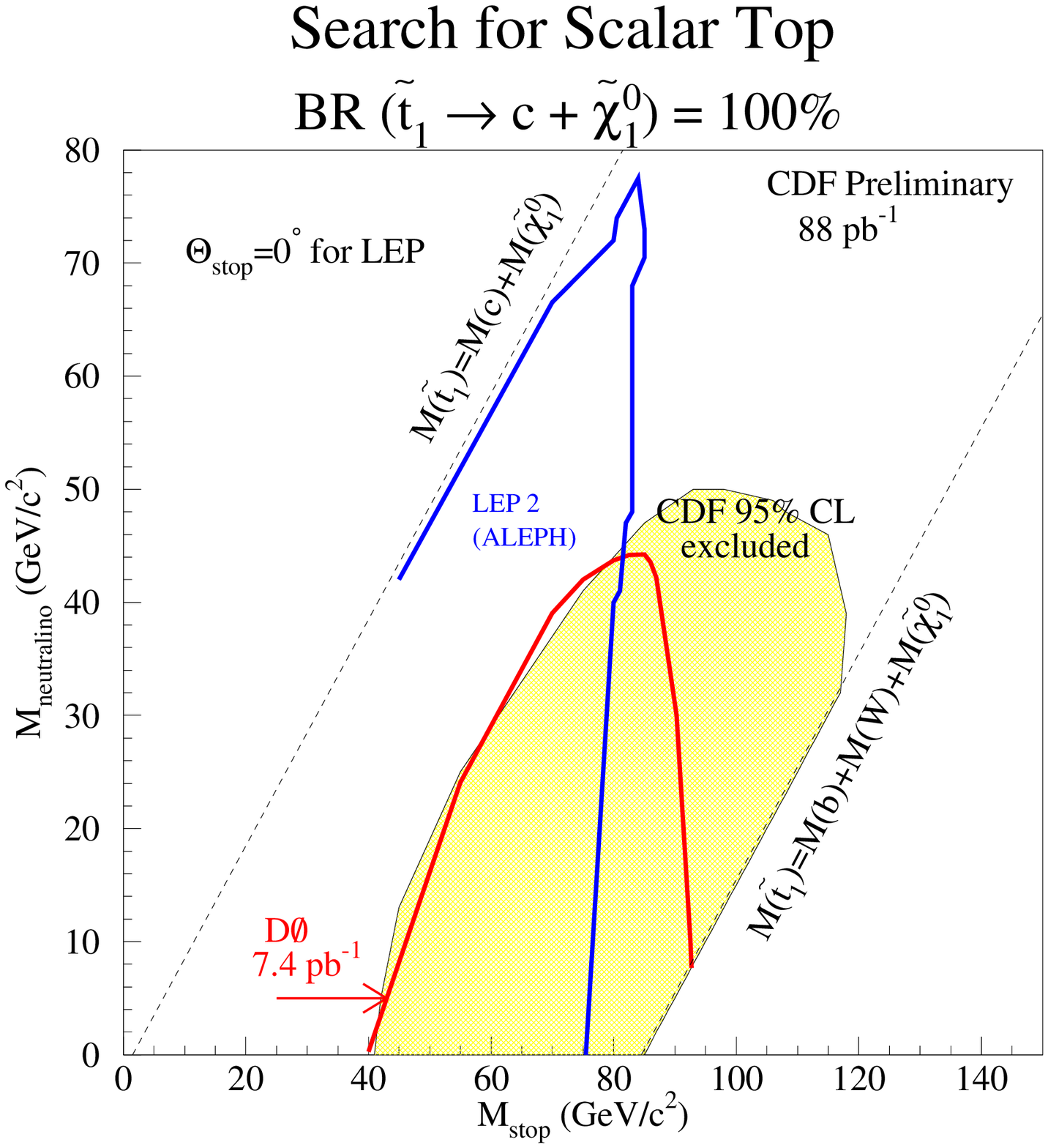}
\vspace{-1.47cm}
\caption{\it 95\% C.L. limit as function of $M_{\tilde{\chi}_{1}^{0}}$ 
and $M_{\tilde{t}_{1}}$, for direct 
$\bar{\tilde{t}}_{1}\tilde{t}_{1}$ production, assuming 
${\mathcal{BR}}(\tilde{t}_{1}\rightarrow c\tilde{\chi}_{1}^{0})=\,100\%$
(CDF experiment).}
\label{dirstoplim}
\end{minipage}\hfill
\vspace{-.39cm}
\end{figure}

\subsection{Search for direct Stop pair production}

Whether $\mathcal{R}$-parity is conserved or not, at Tevatron,  
stop quarks are produced in pairs via $g g$ and 
$q \bar{q}$ fusion.
The LO diagonal pair production cross section 
depends mainly on stop mass and very little 
on other SUSY parameters such as gluino mass, 
\newpage
\begin{center}
\begin{figure}[t!]
\begin{minipage}{2.5in}
  \epsfxsize2.5in 
  \hspace*{0.0cm}\epsffile{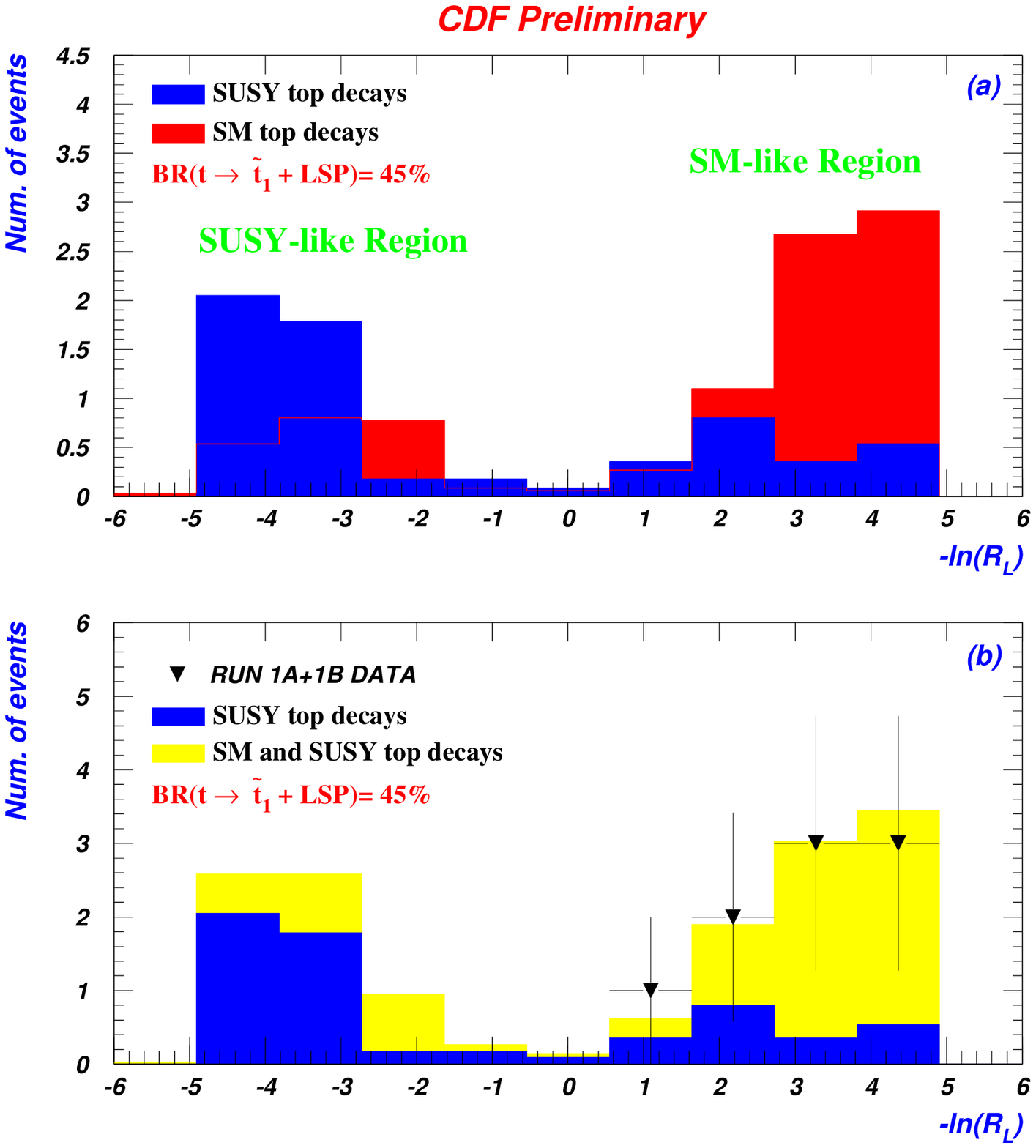}
  \epsfxsize2.5in 
  \hspace*{0.0cm}\epsffile{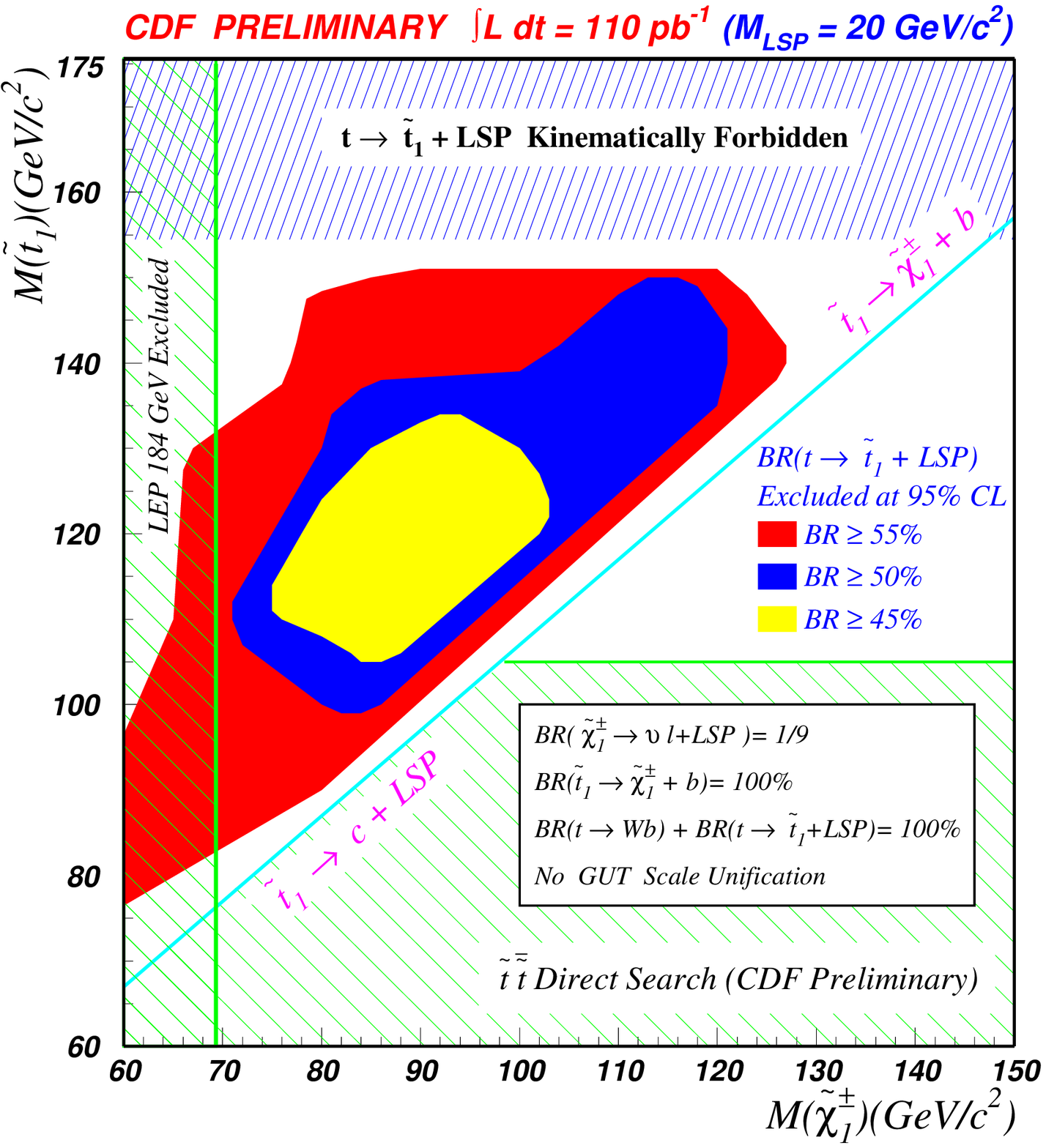}
\end{minipage}\hfill
\end{figure}
\begin{figure}[th!]
\vspace{-0.7cm}
\begin{minipage}{2.4in}
\vspace{-0.4cm}
  \caption{\it  a) Comparison of $-ln(R_{\mathcal{L}})$ for SM top decay 
and SUSY top decays after all the cuts have been applied.
b)  $-ln(R_{\mathcal{L}})$  distribution for Run 1 data (CDF experiment).}
\label{topstop}
\end{minipage}\hfill
\hspace{0.2cm}
\begin{minipage}{2.4in}
\vspace{-0.6cm}
  \caption{\it 95\% C.L. excluded 
            $\mathcal{BR}$($t \rightarrow \tilde{t}_{1} + {\rm LSP}$)
            as function of $m_{\tilde{\chi}^{\pm}_{1}}$ vs $m_{\tilde{t}_{1}}$,
            for $m_{\rm{LSP}}=$ $20$ $GeV/c^{2}$ (CDF experiment).}
\label{carlim20}
\end{minipage}\hfill
\vspace{-.39cm}
\end{figure}
\end{center}
\vspace{-0.25cm}
the masses of the light flavor squarks and the 
mixing angle~\cite{DIRECTSTOP}. 
Whenever kinematically allowed, stop decays into $b\tilde{\chi}^{+}_{1}$;
if this channel is closed, but sneutrino is light, 
then the decay 
$\tilde{t}_{1} \rightarrow b\ell\tilde{\nu}$, dominates.
When both these channels are kinematically suppressed, 
stop decays, via one-loop diagram, to 
charm neutralino ($\tilde{t} \rightarrow c\tilde{\chi}^{0}_{1}$).
The signature for this process is two acollinear 
jets, coming from charmed quarks, a significant 
missing transverse energy contribution 
and no high-$p_{T}$ leptons in the final state.
The events have been selected requiring 2 or 3 jets 
having $E_{T}\ge\,15$ GeV ($|\eta|\le\,2$) and
 $\not\!\!\!E_{\rm T}>40$ $GeV$.
A lifetime tagging technique have been developed for 
$c$-quark identification. 
The probability, that an ensemble 
of tracks in a jet is consistent with being from the 
primary vertex, have been estimated; a probability of less than 5\%, 
at least for one jet, have been required.
The dominant background for this search comes
from $W/Z+$jets, with an unidentified
lepton ($\mu/e$), produced in the $W/Z$ boson decay, 
or from $W\rightarrow \tau\nu_{\tau}$, with 
$\tau$ decaying hadronically.
We observe 11 events, when $14.5 \pm 4.2$ are
expected from SM background (see figure~\ref{dirstop}).
We therefore set a 95\% Confidence Level (C.L.) limit 
as function of the 
$\tilde{t}_{1}$ and $\tilde{\chi}^{0}_{1}$  
masses~\cite{STOPD0} (see figure~\ref{dirstoplim}).

\subsection{Search for SUSY decays of the top}

In the presence of a light stop, $M_{\tilde{t}_{1}}<M_{t}$, 
the top-quark could decay, 
with appreciable branching ratio, into stop
plus neutralino: 
$t \rightarrow \tilde{t}_{1}\tilde{\chi}^{0}_{1}$,
where 
$\tilde{\chi}^{0}_{1}$ is the LSP~\cite{SENDER}.
CDF has searched for such supersymmetric decays of the top, 
in  the kinematic region where stop decays into chargino
plus $b$-quark are dominant 
($M_{\tilde{t}_{1}}>$ $M_{\tilde{\chi}^{\pm}_{1}}$).
The search has been performed on  
the full Run 1 data sample of 
$109.4 \, \pm 7.2 \, {\rm pb}^{-1}$.
The  kinematic differences  between  
SUSY and SM $t\bar{t}$ decays have been used 
to separate signal from background. 
A sample of $\,\,W+\ge3\,\,$ jets top candidate events has been 
\newpage
\begin{figure}[t!]
\begin{minipage}{2.33in}
  \epsfxsize2.45in 
\vspace{-0.5cm}
  \hspace*{0.0cm}\epsffile{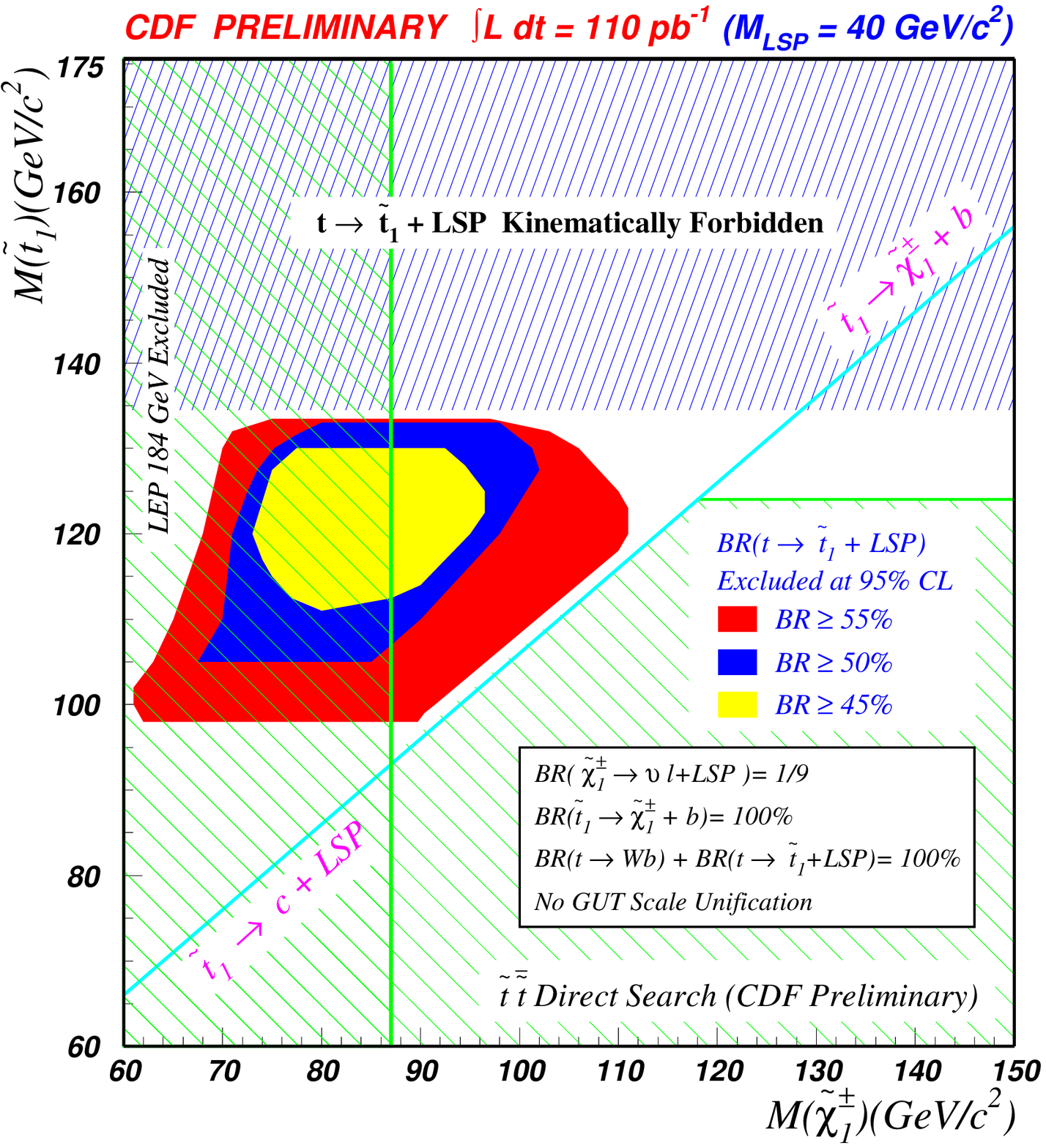}
\vspace{-0.0cm}
  \caption{\it 95\% C.L. excluded 
            $\mathcal{BR}$($t \rightarrow \tilde{t}_{1} + {\rm LSP}$)
            as function of $m_{\tilde{\chi}^{\pm}_{1}}$ vs $m_{\tilde{t}_{1}}$,
            for $m_{\rm{LSP}}=$ $40$ $GeV/c^{2}$ (CDF experiment).}
\label{carlim40}
\end{minipage}\hfill
\begin{minipage}{2.33in}
  \epsfxsize2.3in 
\vspace{-1.1cm}
  \hspace*{0.0cm}\epsffile{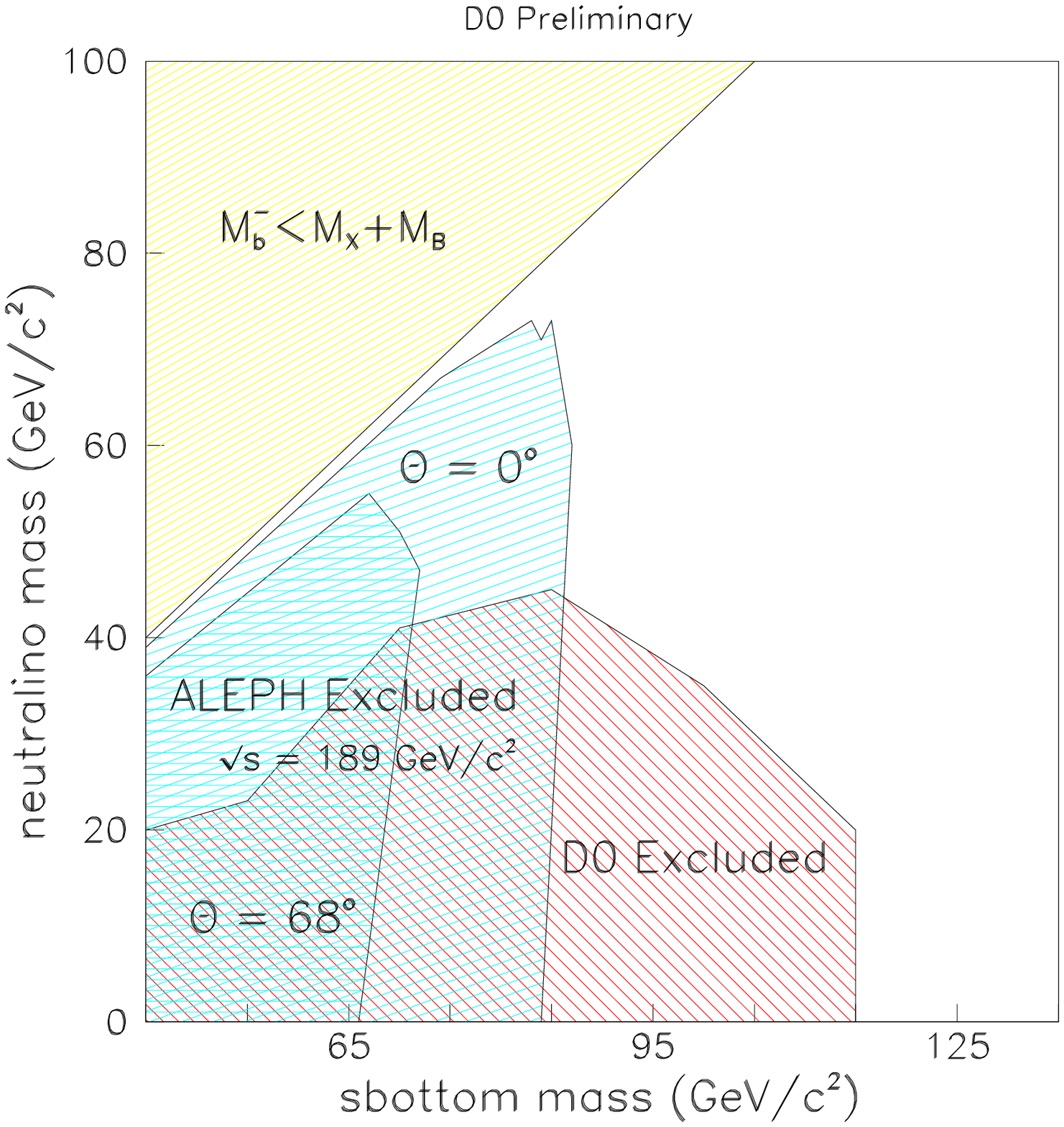}
\vspace{-0.9cm}
\caption{\it 95\% C.L. excluded region in 
the $M_{\tilde{b}_{1}}$, $M_{LSP}$ plane.
The ALEPH contours correspond to 
$\theta=\,0^{\circ}$ and  
$68^{\circ}$ (D$\not$O detector).}
\label{sbottom}
\end{minipage}\hfill
\vspace{-.39cm}
\end{figure}
\noindent
selected, from
the inclusive lepton sample, by requiring:
$E_{\rm T}^{e} >$ $20$ $GeV/c^{2}$, 
$p_{\rm T}^{\mu} >$ $20$ $GeV/c^{2}$, 
$\not\!\!\!E_{\rm T}>$ $25$ $GeV$ and 
the mass of the transverse component of the 
lepton-$\not\!\!\!E_{\rm T}$ system ($M_{T}$) to be 
larger than $40$ $GeV$.
We required the presence of at least 3 jets:
$E^{jet}_{T}(1)>$ $20$ $GeV$,
$E^{jet}_{T}(2)>$ $20$ $GeV$,
$E^{jet}_{T}(3)>$ $15$ $GeV$ ($|\eta_{jet}|<$ $2$),
satisfying the condition that the cosine of the angle, 
in the rest frame, between the proton beam and the jet 
($\theta ^{*}_{i}$), have to be: 
$| cos(\theta_{\rm i} ^{\ast})| _{\rm 1} <$ 0.9,
$| cos(\theta_{\rm j} ^{\ast})| _{\rm 2} <$ 0.8,
$| cos(\theta_{\rm k} ^{\ast})| _{\rm 3} <$ 0.7,
where $| cos(\theta_{\rm j} ^{\ast})| _{\rm k}$ 
are ordered quantities~\cite{topkin}.
Further cuts have been applied to reduce QCD W+jets
and SM top background, by requiring that the three highest
jets are well separated,
the reconstructed transverse momentum of the W is
$p_{T}(W)>$ $50$ $GeV/c$ and
$\not\!\!\!E_{\rm T}>$ $45$ $GeV$.
Finally we required the presence of 
at least one $b$-jet by asking a SVX $b$-tag in the event.
The discrimination between SUSY top decays and 
SM top background has been achieved by combining 
the information on the $E_{T}^{jet}$
in a Relative Likelihood  (see figure~\ref{topstop}), defined as:
$R_{\mathcal{L}}= \frac{{\mathcal{L}}_{Abs}^{SUSY}}{{\mathcal{L}}_{Abs}^{SM}}$,
where ${\mathcal{L}}_{Abs}$ is the
Absolute  Likelihood function defined by the equation:
${\mathcal{L}}_{Abs}= \left( \frac{1}{\sigma} \frac{d\sigma}{dE_{T}^{jet}(2)}    \right)\times
\left( \frac{1}{\sigma} \frac{d\sigma}{dE_{T}^{jet}(3)}  \right)$.
Since no signal has been observed, we set a 95\%
{\it C.L.} limit on 
${\mathcal{BR}} ( t \rightarrow \tilde{t}_{1} \tilde{\chi}^{0}_{1})$, 
as function of $M_{\tilde{t}_{1}}$ and $M_{\tilde{\chi}^{\pm}_{1}}$, 
for $M_{LSP}$ between $20$ and $40$ $GeV/c^{2}$
(see figure~\ref{carlim20} and~\ref{carlim40}).

\section{Search for scalar bottom}

At large tan$\beta$ (tan$\beta>$ $10$)~\cite{SBOTTOM}, 
a considerable $\tilde{b}_{L}-\tilde{b}_{R}$ mixing 
can occur in the sbottom sector, 
leading to a scenario in which the 
$\tilde{b}_{1}$ could be the lightest scalar quark.
The LO and NLO diagonal pair production cross section
are the same as for stop quark~\cite{DIRECTSTOP}.
D$\not$O  has performed a search for light sbottom, 
assuming $\tilde{\chi}^{0}_{1}$ to be the LSP 
and fixing the branching ratio
${\mathcal{BR}}(\tilde{b}_{1}\rightarrow b\tilde{\chi}_{1}^{0})=\,100\%$.
In the region of interest for the Teva-
\newpage
\begin{center}
\begin{figure}[h!]
\begin{minipage}{5.0in}
  \epsfxsize2.4in 
  \hspace*{-0.2cm}\epsffile{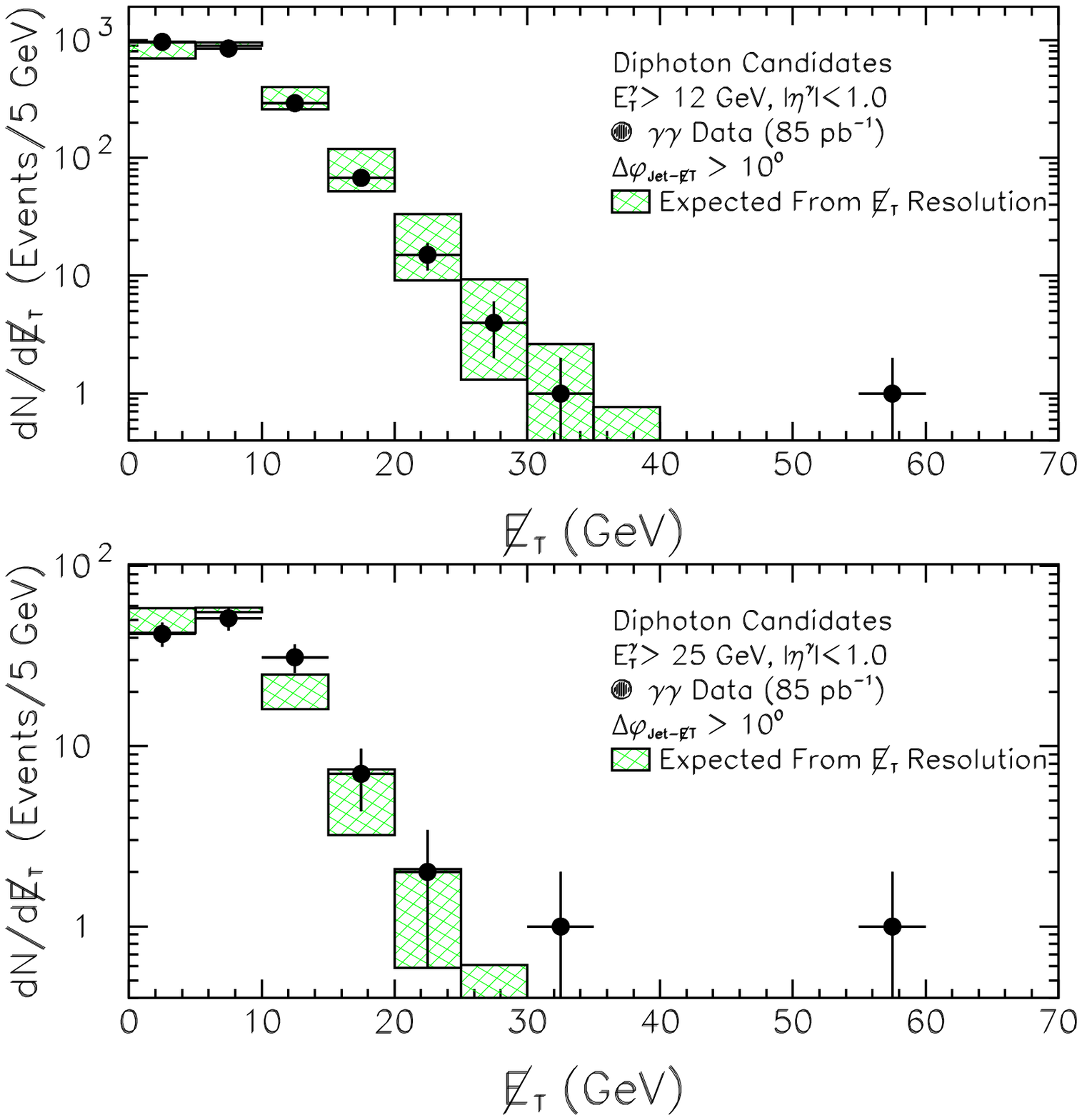}
  \epsfxsize2.4in 
  \hspace*{0.5cm}\epsffile{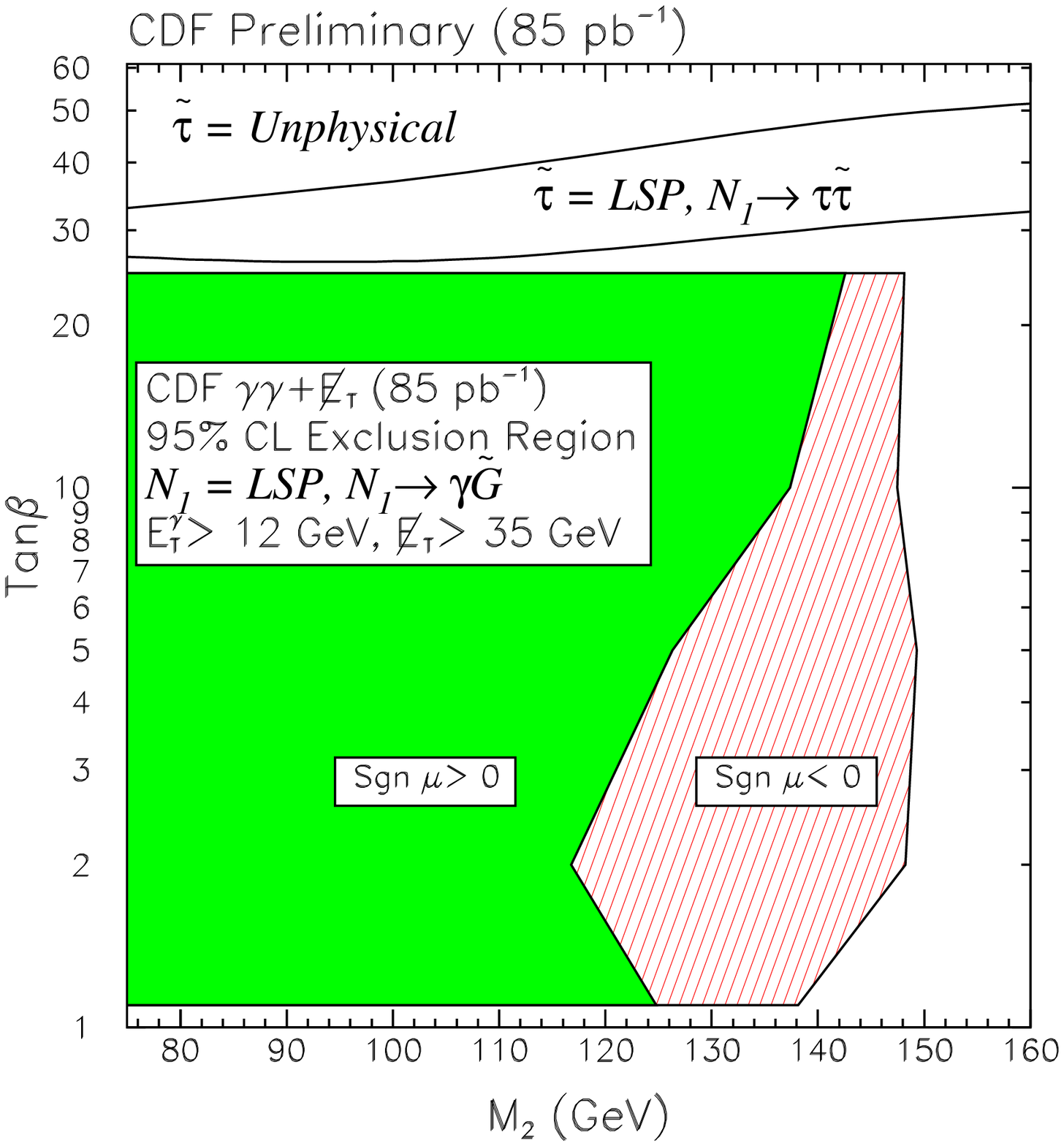}
\end{minipage}\hfill
\end{figure}
\begin{figure}[th!]
\vspace{-1.7cm}
\begin{minipage}{2.4in}
  \caption{\it The  $\not\!\!\!E_{\rm T}$ spectrum for diphoton events
with $E^{\gamma}_{T}>12\,GeV$ and with  $E^{\gamma}_{T}>25\,GeV$ in the
data from CDF detector.}
\label{LGLSP}
\end{minipage}\hfill
\hspace{0.2cm}
\begin{minipage}{2.4in}
\vspace{0.0cm}
  \caption{\it 95\% C.L. excluded region ($M_{2}$ vs $tan\beta$)
for the Light Gravitino LSP scenario (CDF experiment).}
\label{LGLSPlim}
\end{minipage}\hfill
\vspace{-.39cm}
\end{figure}
\end{center}
tron,  
the sbottom can only decay into $ b\tilde{\chi}_{1}^{0}$. 
The decays of the sbottom into 
$\tilde{t}_{1}W$, with a virtual W,
$t\tilde{\chi}_{1}^{-}$, or into
$\tilde{t}_{1}H^{\pm}$ are kinematically suppressed.
The  signature for the process under study is therefore
2 $b$-quarks plus  $\,\not\!\!\!E_{\rm T}$. 
This final state is similar to 
those of two published D$\not$O analysis:
the search for third generation leptoquark
($\nu\nu bb$)~\cite{D0leptoquark},
and the search for direct scalar top production
($\bar{\tilde{t}_{1}}\tilde{t}_{1}\rightarrow 
c\bar{c}\tilde{\chi}^{0}_{1}\tilde{\chi}^{0}_{1}$)~\cite{D0stop}.
The limits on sbottom production have been obtained 
combining these two samples.
The number of events observed in the data is 5;
$6.0\pm1.3$ events are expected from SM background.
Figure~\ref{sbottom} plots the 95\% C.L. limit, as function 
of $M_{\tilde{b}_{1}}$ and $M_{\tilde{\chi}_{1}^{0}}$.
A lower limit on the sbottom mass of $M_{\tilde{b}}>115$ $GeV/c^{2}$
which is valid for $M_{\tilde{\chi}_{1}^{0}}<20$  $GeV/c^{2}$ 
has been placed.

\section{Photon enriched SUSY}

In the last few years, triggered by the CDF 
$ee\gamma\gamma+\not\!\!\!E_{\rm T}$ 
event candidate~\cite{eeggmet}, many models of new physics, 
predicting photon enriched final states 
have been introduced.
CDF and D$\not$O have systematically searched for events 
containing two photons in the final state: $\gamma\gamma+X$,
where X can be a jet, a b-tag, a lepton
($e$, $\mu$, $\tau$) or $\not\!\!\!E_{\rm T}$.

\subsection{Light Gravitino LSP}

In the Minimal Gauge Mediated SUSY Breaking Model (MGM), 
as in SUGRA, the supersymmetry breaking occurs in a hidden sector.
Unlike the SUGRA models, in which $\Lambda_{SUSY}\sim 10^{11}$,
in MGM  $\Lambda_{SUSY}\sim 10^{5}-10^{9}$, making the Gravitino
the lightest supersymmetric particle ($M_{\tilde{G}}\sim$ $1 \div 10^{2}eV$).
 \hspace{0.2cm}The Next-to-lightest \hspace{0.1cm}SUSY 
\newpage
\begin{center}
\begin{figure}[h!]
\begin{minipage}{5.0in}
  \epsfxsize2.4in 
  \hspace*{-0.2cm}\epsffile{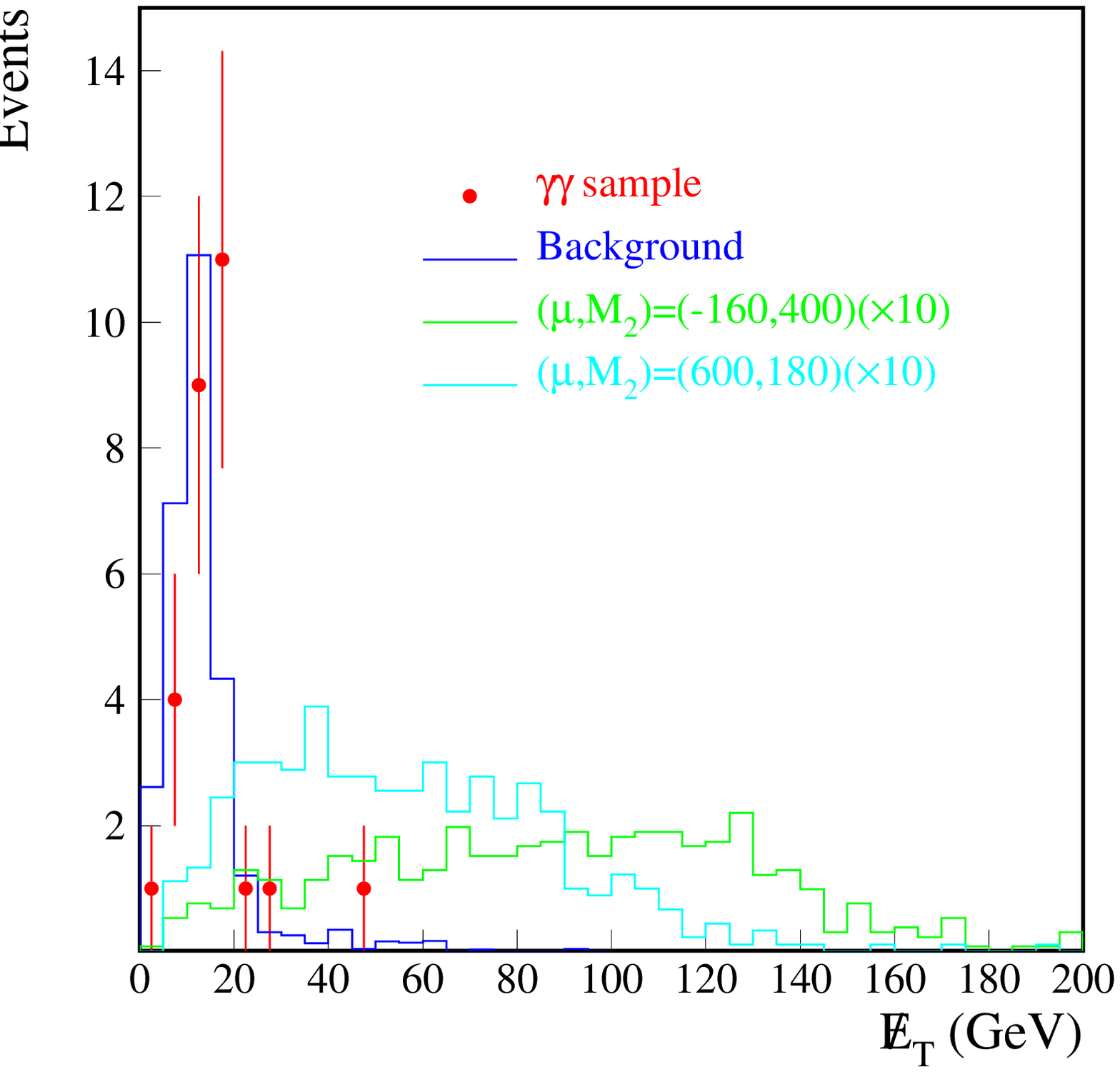}
  \epsfxsize2.4in 
  \hspace*{0.5cm}\epsffile{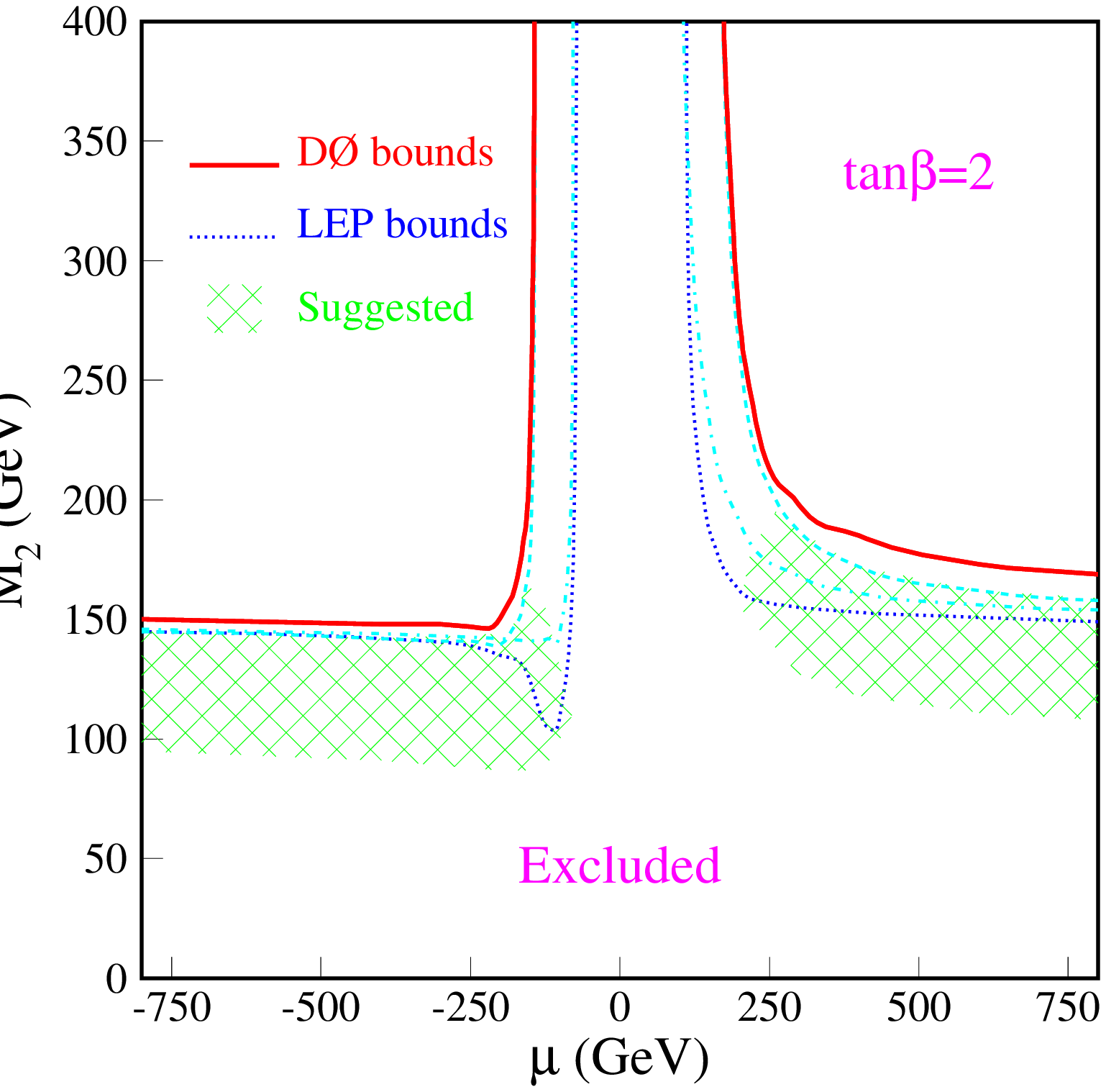}
\end{minipage}\hfill
\end{figure}
\begin{figure}[th!]
\vspace{-1.0cm}
\begin{minipage}{2.4in}
\vspace{-0.37cm}
  \caption{\it The  $\not\!\!\!E_{\rm T}$ spectrum for diphoton events
with $E^{\gamma}_{T}(1)>20\,GeV$ and  $E^{\gamma}_{T}(2)>12\,GeV$ in the
data from D$\not$O experiment.}
\label{LGLSPD0}
\end{minipage}\hfill
\hspace{0.2cm}
\begin{minipage}{2.4in}
\vspace{0.0cm}
  \caption{\it  95\% C.L. excluded region ($\mu$ vs $M_{2}$)
for the Light Gravitino LSP scenario (D$\not$O detector);
the dashed area shows the region suggested to explain the 
CDF $ee\gamma\gamma+\not\!\!\!E_{\rm T}$ 
event.}
\label{LGLSPlimD0}
\end{minipage}\hfill
\vspace{-.39cm}
\end{figure}
\end{center}
particle (NLSP), which in the models is
the neutralino ($\tilde{\chi}_{1}^{0}$), will decay dominantly via
$\tilde{\chi}_{1}^{0}\rightarrow \gamma\tilde{G}$.
Therefore, any pair produced sparticle will produce two photons 
and $\not\!\!\!E_{\rm T}$~\cite{GORDY}.
In this model the  $\tilde{\chi}_{1}^{0}$ 
lifetime depends on $M_{\tilde{G}}$ and is assumed to be less
than 1 KeV. The gaugino masses are defined by $M_{2}$, tan$\beta$ and
sign$\mu$ and $M_{\tilde{\chi}^{\pm}_{1}}\sim M_{\tilde{\chi}_{1}^{0}}\sim M_{2}$. 
In order to set a limit in this MGM model, the results of the 
counting experiment in the $\gamma\gamma\not\!\!\!E_{\rm T}$
have been used. 
Events are asked to have 2 photons with 
$E^{\gamma}_{T}>$ $12$ $GeV$ ($|\eta|<$ $1$)
and  $\not\!\!\!E_{\rm T}>35$
or 
$E^{\gamma}_{T}>$ $25$ $GeV$ ($|\eta|<$ $1$)
and  $\not\!\!\!E_{\rm T}>25$.
For  $\,\not\!\!\!E_{\rm T}>35$ $GeV$, 1 event 
passes the cuts, when the expected background is 
$0.5\pm0.1$ (see figure~\ref{LGLSP})~\cite{eeggmet}.
Figure~\ref{LGLSPlim} shows the contour plot of the 
95\% C.L. excluded region as function of 
tan$\beta$ and $M_{2}$.

\noindent
An analogous search has been performed by 
D$\not$O Collaboration by selecting events with 
$E^{\gamma1}_{T}>$ $20$ $GeV$ and
$E^{\gamma2}_{T}>$ $12$ $GeV$ ($|\eta|<$ $1.2$ or 
$1.5<|\eta|<2.0$).
For  $\not\!\!\!E_{\rm T}>25$ $GeV$, 
2 events are observed when $2.3\pm0.9$ are expected
from SM background processes (see figure~\ref{LGLSPD0}).
Figure~\ref{LGLSPlimD0} plots
the 95\% C.L. excluded region ($\mu$ vs $M_{2}$).

\subsection{Neutralino Radiative Decay}

The $\gamma\not\!\!\!E_{\rm T}+n$ jets event topology may also 
arise in MSSM, in some region of the parameter space,
where the radiative decay 
$\tilde{\chi}^{0}_{2} \rightarrow \gamma \tilde{\chi}^{0}_{1}$
dominates. Depending on the number of $\tilde{\chi}^{0}_{2}$,
involved in the process, it will be possible to have one or 
more $\gamma$ in the final state. 
D$\not$O has searched for  
such signal using 99 $pb^{-1}$ of data,  
assuming that slepton masses are heavy, 
$M_{\tilde{\chi}^{0}_{2}}-M_{\tilde{\chi}^{0}_{1}}>$ $20$ $GeV/c^{2}$ 
and ${\mathcal{BR}}(\tilde{\chi}^{0}_{2}
\rightarrow \gamma\tilde{\chi}^{0}_{1})=$ $100$\%.
In order to select  $\gamma\not\!\!\!E_{\rm T}+\ge2$ jets 
event candidates, we require to have at least one 
identified photon with $E^{\gamma}_{T}>$ $20$ $GeV$, 
$\not\!\!\!E_{\rm T}>$ $25$  $GeV$ (see fig.~\ref{raddec})
and two or more jets ($E^{jet}_{T}>$ $20$ $GeV$).
Then the event selection has been optimized 
in the $\not\!\!\!E_{\rm T}$, $H_{T}$ plane 
($\not\!\!\!E_{\rm T}>$ $45$ $GeV$, $H_{T}>$ $220$ $GeV$), 
resulting in 5 events passing 
the above cuts. $8.1\pm5.8$ events are expected 
from SM background processes.
The 95\% C.L. upper limit on  $\sigma\times{\mathcal{BR}}$
as function of $M_{\tilde{q}}$ ($M_{\tilde{q}}=M_{\tilde{g}}$) 
is given in figure~\ref{raddeclim}.
\newpage
\begin{center}
\begin{figure}[h!]
\begin{minipage}{5.0in}
  \epsfxsize2.4in 
  \hspace*{-0.2cm}\epsffile{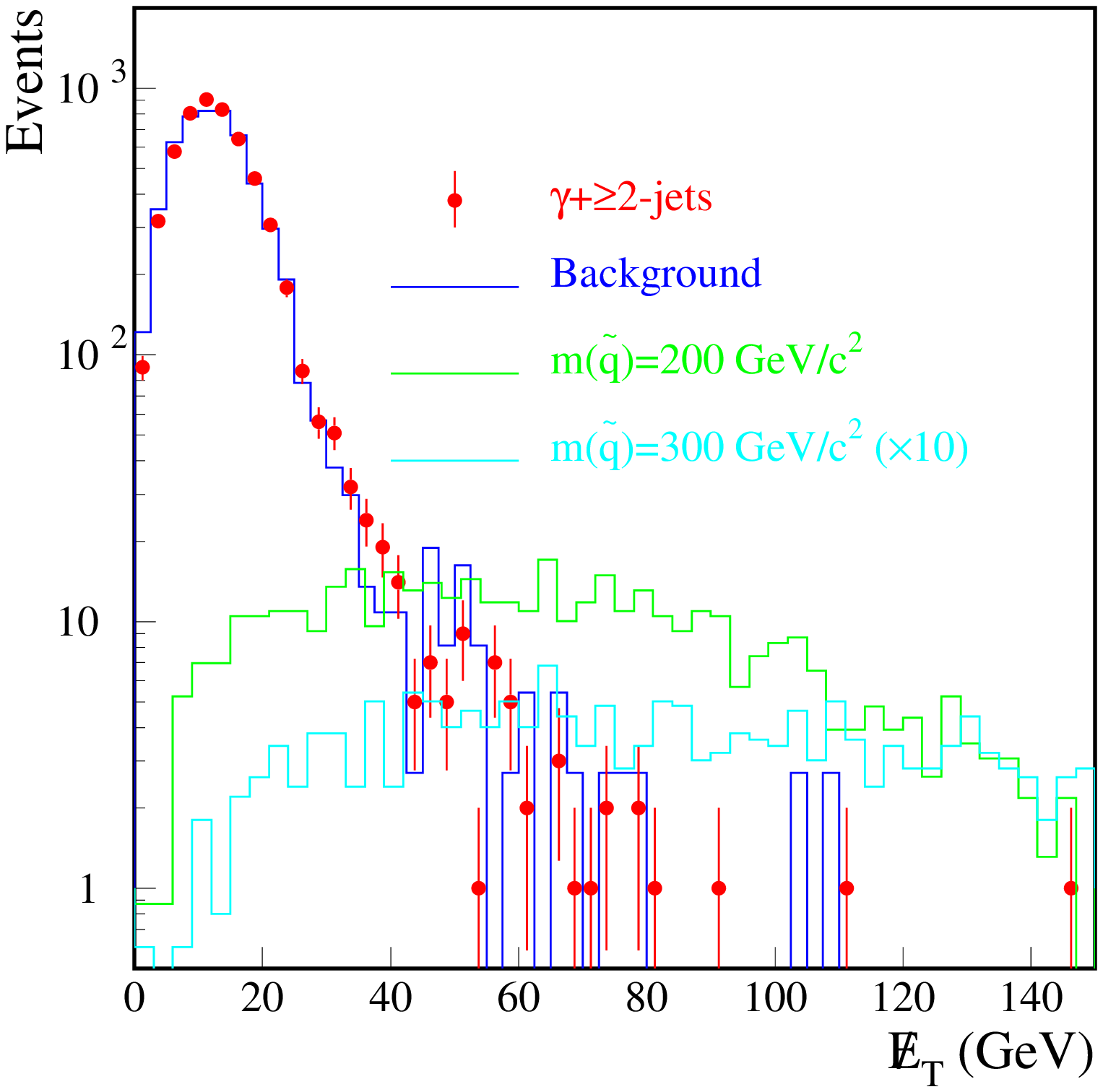}
  \epsfxsize2.4in 
  \hspace*{0.2cm}\epsffile{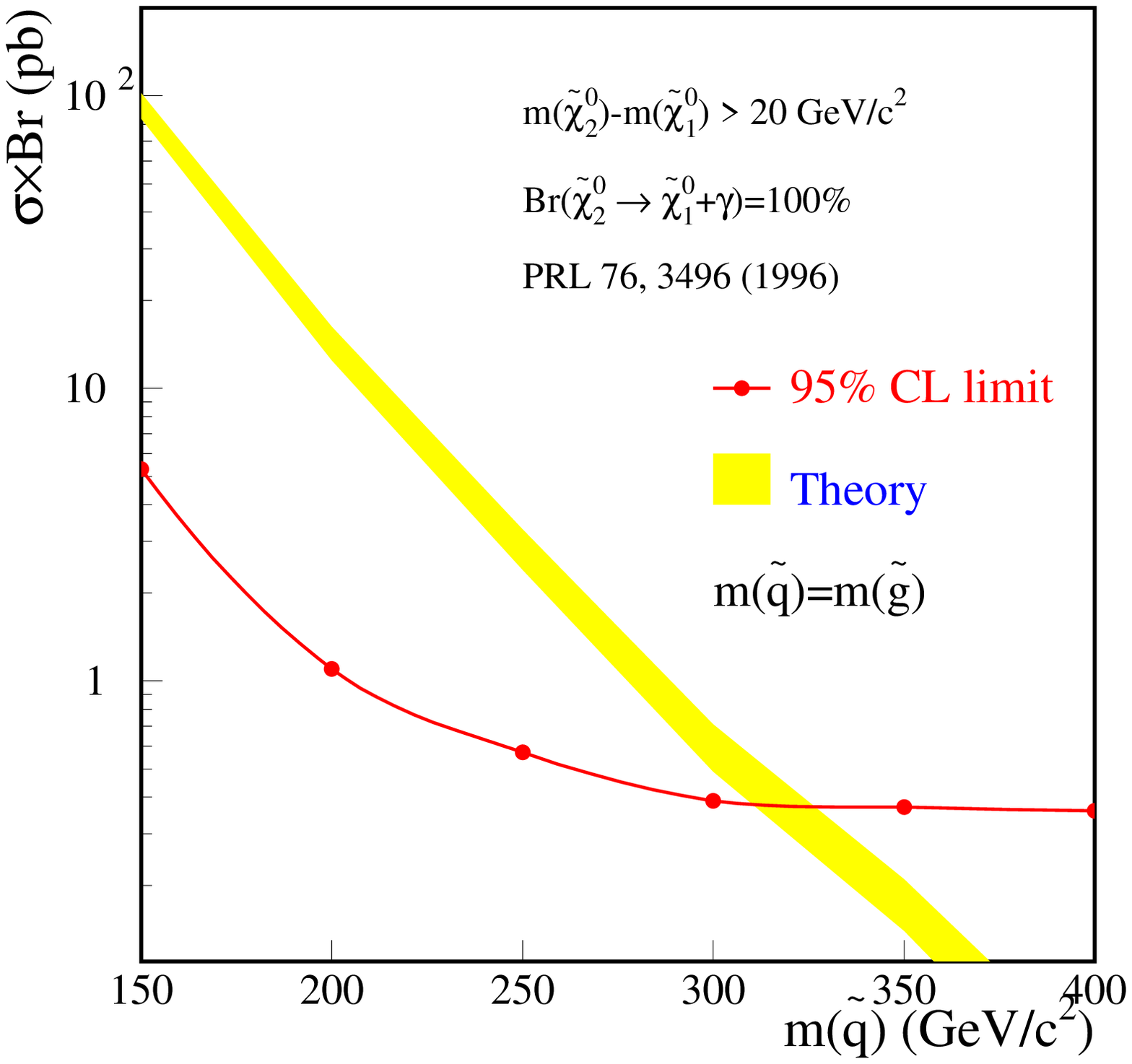}
\end{minipage}\hfill
\end{figure}
\begin{figure}[th!]
\vspace{-0.7cm}
\begin{minipage}{2.4in}
\vspace{-0.37cm}
  \caption{\it The  $\not\!\!\!E_{\rm T}$ spectrum for events 
with one photon ($E^{\gamma}_{T}(1)>20\,GeV$) and  two or more jets 
($E^{jet}_{T}>20\,GeV$) in the data from D$\not$O experiment.}
\label{raddec}
\end{minipage}\hfill
\hspace{0.2cm}
\begin{minipage}{2.4in}
\vspace{0.0cm}
  \caption{\it The 95\% C.L. upper limit on $\sigma\times{\mathcal{BR}}$ 
as function of $M_{\tilde{q}}=M_{\tilde{g}}$. 
The hatched band represents the range of theoretical cross 
section for different sets of MSSM parameter values (D$\not$O experiment).}
\label{raddeclim}
\end{minipage}\hfill
\vspace{-.8cm}
\end{figure}
\end{center}
\subsection{The Higgsino LSP scenario}

The present Higgsino LSP scenario 
assumes MSSM without sfermion/scalar 
masses unification.
$\tilde{\chi}^{0}_{1}$ is an Higgsino-like LSP
(tan$\beta=1.2$), and  
$\tilde{\chi}^{0}_{2}$ is photino-like ($M_{1}=M_{2}$).
So doing this model predicts a large branching ratio for the process
$\tilde{\chi}^{0}_{2}\rightarrow \gamma\tilde{\chi}^{0}_{1}$
and a light stop. A branching ratio of
${\mathcal{BR}}(\tilde{t}_{1}\rightarrow c\tilde{\chi}^{0}_{1})=100\%$
is assumed. We end up with a topology containing
$\gamma b + \not\!\!\!E_{\rm T}$ in the final state:
$p \bar{p} \rightarrow C_{1}\,\,N_{2} \rightarrow \tilde{t}\,b\,\gamma\,N_{1}\rightarrow c\,b\,\gamma\,N_{1}\,N_{1}$.
CDF has searched for such events 
requiring a photon with $E^{\gamma}_{T}>25$ $GeV$, a
SVX $b$-tag and $\not\!\!\!E_{\rm T}>25$ $GeV$.
To increase the sensitivity we then required
the photon to be not opposite to the $\not\!\!\!E_{\rm T}$,
and $\,\,\not\!\!\!E_{\rm T}>40$ $GeV$ (see figure~\ref{higgsino1}).
We see 2 events; this allows to 
rule out more than approximately
7 events of anomalous production.
Figure~\ref{higgsino2}
shows the limit, plotted as function of gluino mass. 
In figure~\ref{higgsino3} we present 
the limit on direct production of 
$\tilde{\chi}^{\pm}_{2}\tilde{\chi}^{0}_{2}$.

\subsection{Gauge-mediated SUSY with $\gamma b+\not\!\!\!E_{\rm T}$}

Another way to produce the $\gamma b+\not\!\!\!E_{\rm T}$ 
signature comes from a GMSB model. 
In this model, the gravitino is light and becomes the
LSP. The $\tilde{\chi}^{0}_{1}$ 
is an higgsino and may decay into either a 
gravitino and a photon 
($\tilde{\chi}^{0}_{1} \rightarrow \tilde{G}\gamma$) or 
into a gravitino and an Higgs boson 
($\tilde{\chi}^{0}_{1} \rightarrow \tilde{G}h$), 
with the Higgs decaying into $b\bar{b}$. 
Since we have two $\tilde{\chi}^{0}_{1}$ 
in each event this will give rise to
a $\gamma b+\not\!\!\!E_{\rm T}$ signature.
The CDF limits obtained in this scenario
are shown in figure~\ref{GMSB}.


\section{Conclusions}
Extensive searches of supersymmetric signals have been 
done at Tevatron Collider. 
No positive results have been found so far
showing that the data are consistent with 
\newpage
\begin{center}
\begin{figure}[h!]
\begin{minipage}{1.5in}
  \epsfxsize2.4in 
  \hspace*{0.0cm}\epsffile{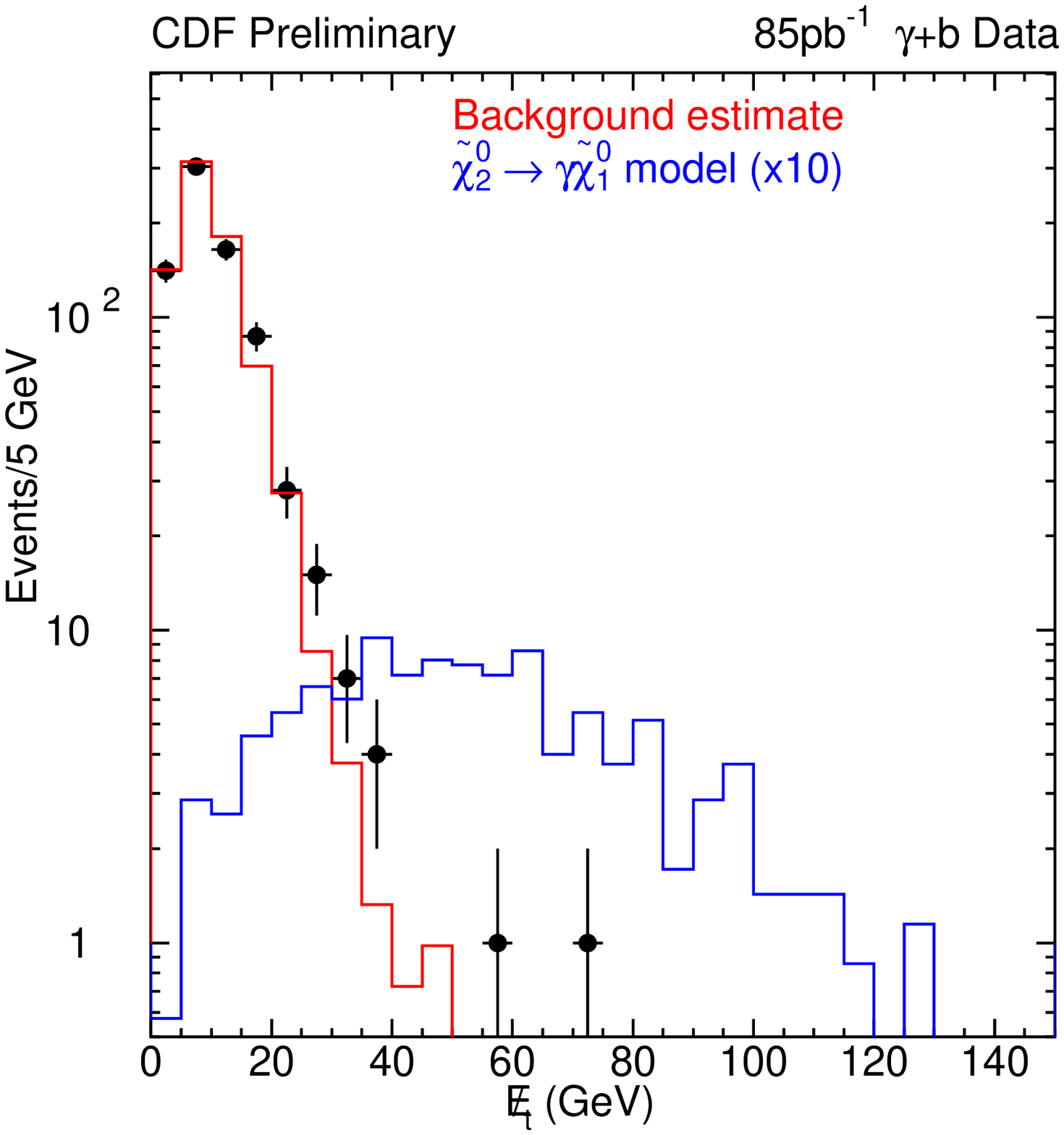}
\end{minipage}\hfill
\begin{minipage}{1.0in}
  \epsfxsize2.9in 
  \vspace*{-0.7cm}
  \hspace*{-4.45cm}\epsffile{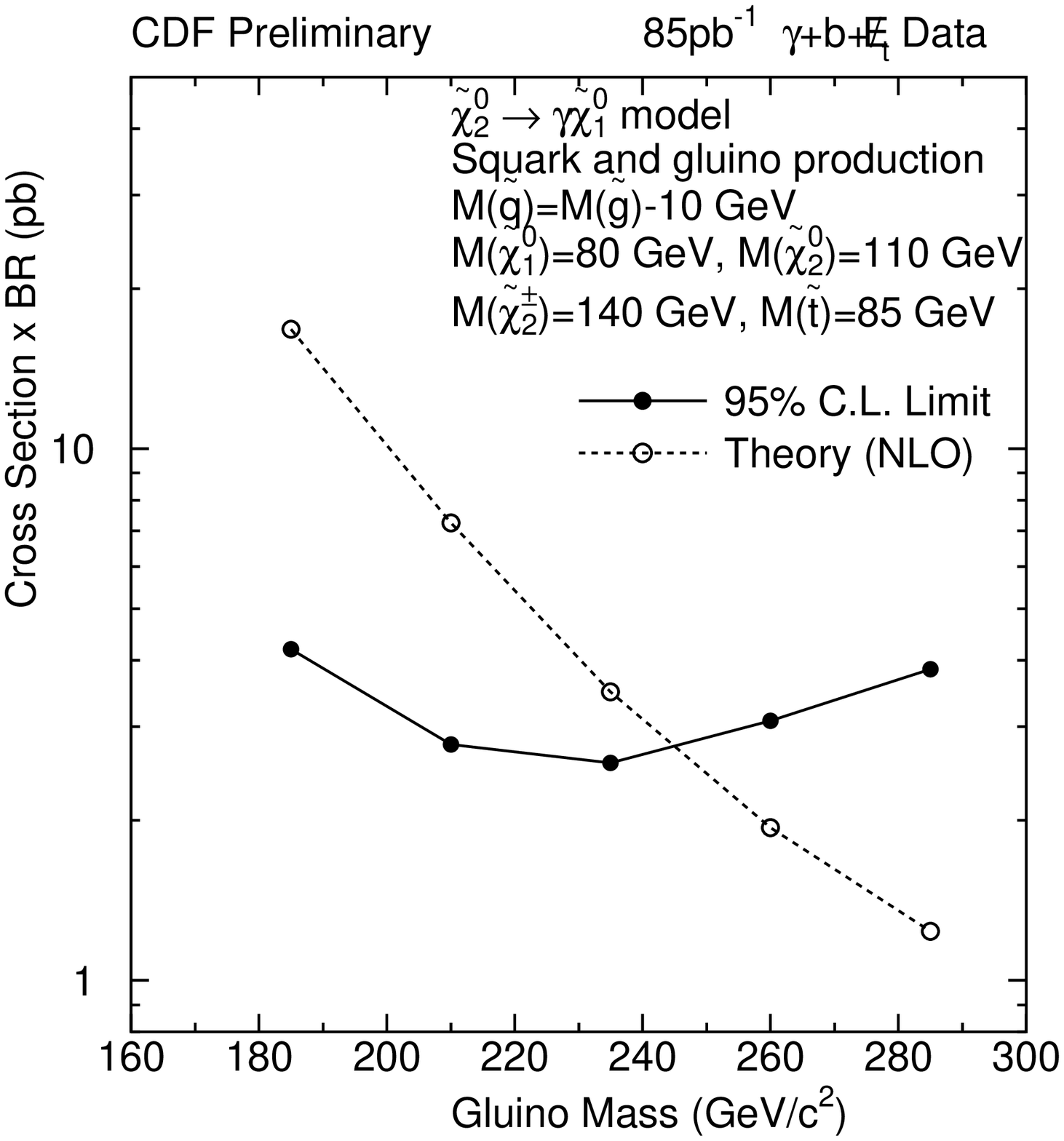}
\end{minipage}\hfill
\end{figure}
\vspace{-0.1cm}
\begin{figure}[th!]
\vspace{-1.7cm}
\begin{minipage}{2.4in}
  \caption{\it The  $\,\,\not\!\!\!E_{\rm T}$ distribution in the data from 
CDF detector, compared to the SUSY Higgsino LSP baseline model.} 
\label{higgsino1}
\end{minipage}\hfill
\hspace{0.2cm}
\begin{minipage}{2.4in}
\vspace{0.0cm}
  \caption{\it The limits on the $\sigma\times{\mathcal{BR}}$ 
for SUSY production in the Higgsino LSP model
(CDF experiment).}
\label{higgsino2}
\end{minipage}\hfill
\vspace{-.39cm}
\end{figure}
\end{center}
the SM expectation. Further regions of SUSY 
parameter space, 
assuming different SUSY models, 
have been excluded.
With the Run II~\cite{UPGRADES} upgrades, providing an higher 
acceptance and higher luminosity, it will be possible
to probe larger region of SUSY parameter space~\cite{RUNIIPHYS}.


\eject


\vspace*{5pt}
\section*{References}
\begin{thebibliography}{99}

\bibitem{Altarelli}
G.~Altarelli, CERN-TH-98-348, {\bf hep-ph/9811456} (1998).

\bibitem{SUSYTEO}S.P.~Martin, {\bf hep-ph/9709356}.

\bibitem{BOER}
W.~de Boer, {\bf hep-ph/9808448}

\bibitem{DREINER}
H.~Dreiner, {\it Pramana} {\bf 51}, 123 (1998).

\bibitem{VISSANI}
J.~Ellis {\it et al.}, 
 \Journal{\PLB}{150}{142}{1985}.\\
F.~Vissani,  IC-96-32, {\bf hep-ph/9602395} (1995).


\bibitem{GMSB1}M. Dine {\it et al.}, 
\Journal{\NPPS}{62}{266-275}{1998}.


\bibitem{GMSB2}G.F. Giudice, R. Rattazzi, CERN-TH-97-380, 
{\bf hep-ph}/9801271 (1998).

C.~Kilda, Nucl. Phys. Proc. Suppl. 62, 485 (1998).


\bibitem{GENSTOP}
Ken-ichi~Hikasa, Makoto~Kobayashi, 
   \Journal{\PRD}{36}{724}{1987}.

\bibitem{DIRECTSTOP}
W.~Beenakker {\it et al.}, 
              \Journal{\NPB}{515}{3}{1997}.
\bibitem{STOPD0}
S.~Abachi {\it et al.}(D$\not$O Collaboration), 
              \Journal{\PRL}{76}{2222}{1996}. 


\bibitem{SENDER}
   J.~Sender,
   \Journal{\PRD}{54}{3271}{1996}.\\
   J.~D. Wells and G. L. Kane,
   \Journal{\PRL}{76}{869}{1996}. \\
   G.~Mahlon and G. L. Kane,  
   \Journal{\PRD}{55}{2779}{1997}. \\
   M.~Hosch {\it et al.},
   \Journal{\PRD}{58}{034002}{1999}.

\newpage
\begin{center}
\begin{figure}[h!]
\begin{minipage}{1.5in}
  \epsfxsize2.9in 
  \vspace*{-0.7cm}
  \hspace*{-1.0cm}\epsffile{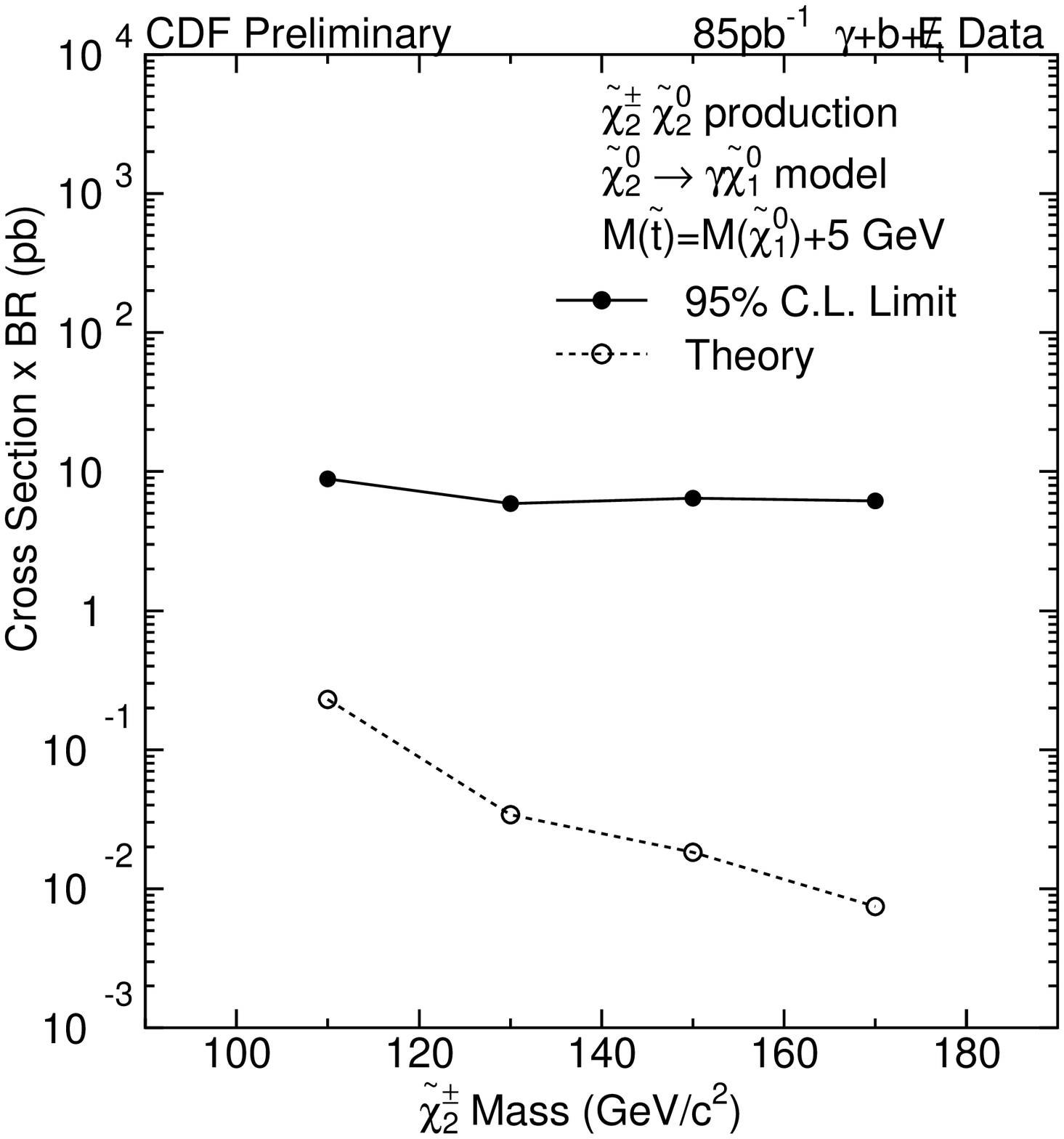}
\end{minipage}\hfill
\begin{minipage}{1.0in}
  \epsfxsize2.9in 
  \vspace*{-0.7cm}
  \hspace*{-4.45cm}\epsffile{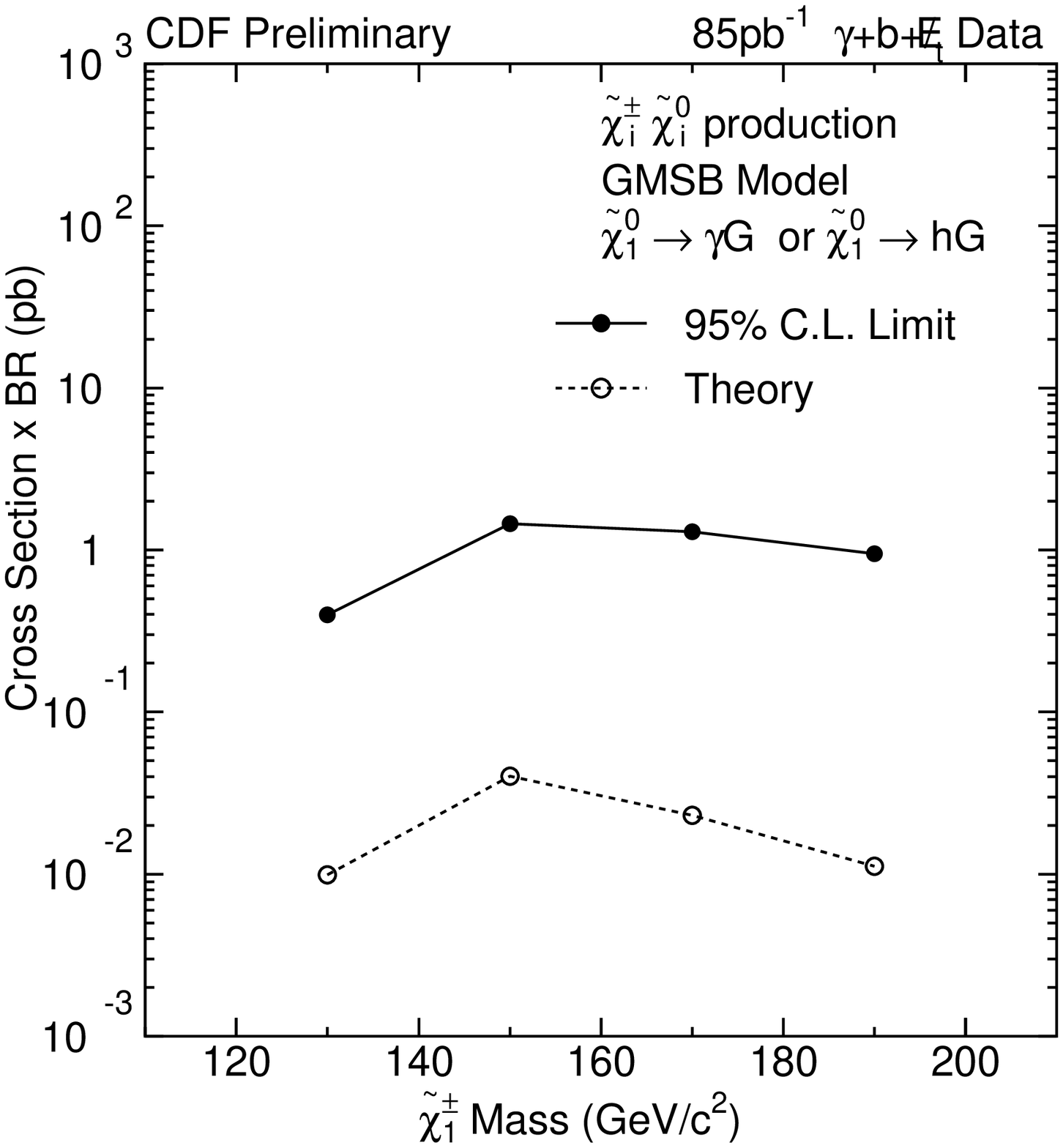}
\end{minipage}\hfill
\end{figure}
\vspace{-6.cm}
\begin{figure}[th!]
\vspace{-2.7cm}
\begin{minipage}{2.4in}
  \caption{\it The limits on the $\tilde{\chi}^{\pm}_{2}\tilde{\chi}^{0}_{2}$
               cross section in the SUSY Higgsino LSP model
               (CDF experiment).}
\label{higgsino3}
\end{minipage}\hfill
\hspace{0.2cm}
\begin{minipage}{2.4in}
\vspace{-0.2cm}
  \caption{\it The limits on the SUSY $\sigma\times{\mathcal{BR}}$ 
in the MGMSB model (CDF experiment).}
\label{GMSB}
\end{minipage}\hfill
\end{figure}
\end{center}
\vspace{-10.2cm}


\bibitem{topkin}
   F.~Abe {\it et al.}(CDF Collaboration),
   \Journal{\PRD}{51}{4623}{1995}.\\
   F.~Abe {\it et al.} (CDF Collaboration),
   \Journal{\PRL}{74}{2626}{1995}. 

\bibitem{SBOTTOM}
  A.~Bartl {\it et al.}, Z. Phys. {\bf C76}, 549 (1997).

\bibitem{D0stop}
  F.~Abachi {\it et al.} (D$\not$O Collaboration),
   \Journal{\PRL}{76}{2222}{1996}. 

\bibitem{D0leptoquark}
  F.~Abachi {\it et al.} (D$\not$O Collaboration),
   \Journal{\PRL}{81}{38}{1998}. 

\bibitem{eeggmet}
  F.~Abe {\it et al.}, (CDF Collaboration),
       \Journal{\PRD}{81}{1971}{1998}.\\
  F.~Abe {\it et al.}, (CDF Collaboration),{\bf hep-ex/9806034}.

\bibitem{GORDY} S.~Ambrosanio {\it et al.},
   \Journal{\PRD}{54}{5395}{1996}. 

\bibitem{UPGRADES}
R.~Blair {\it et al.}, (CDF-II Collaboration), 
FERMILAB-PUB-96-390-E (1996).

\bibitem{RUNIIPHYS}
D.~Amidei {\it et al.}, (TeV-2000 Study Group Collaboration), 

\end{thebibliography}
\end{document}